\documentclass[12pt]{article}%
\usepackage{amsmath,latexsym}
\usepackage{graphicx}
\usepackage{amsmath}
\usepackage{amsfonts}
\usepackage{amssymb}%
\begin{document}

\title{{\Huge A comparison of Ho\v{r}ava-Lifshitz gravity
and Einstein gravity through  thin-shell
wormhole construction}}
\author{{\ F.Rahaman \thanks{farook\_rahaman@yahoo.com}, P.K.F.Kuhfittig
\thanks{kuhfitti@msoe.edu}, M.Kalam
\thanks{mehedikalam@yahoo.co.in}}, A.A.Usmani \thanks{anisul@iucaa.ernet.in}~ and  Saibal Ray
\thanks{saibal@iucaa.ernet.in}
  \and $^{*}$ {\small
Department of Mathematics, Jadavpur University, Kolkata - 700032,
India} \and $^{\dag}$ {\small  Department of Mathematics,
Milwaukee School of Engineering, Milwaukee, Wisconsin 53202-3109,
USA}  \and $^{\ddag}${\small Department of Physics, Aliah
University, Sector - V , Salt Lake,  Kolkata - 700091, India}
 \and $^{\S}${\small Department of Physics,
Aligarh Muslim University, Aligarh 202 002, Uttar Pradesh, India}
 \and $^{\P}${\small Dept.  of Physics,
Govt.  College of Engineering \& Ceramic Technology, Kolkata 700
010, India}} \maketitle
\date{}

\begin{abstract}\noindent
In this paper we have  constructed a new class of thin-shell
wormholes from black holes in Ho\v{r}ava-Lifshitz gravity.
Particular emphasis is placed on those aspects that allow a
comparison of Ho\v{r}ava-Lifshitz to Einstein gravity.  The
former enjoys a number of advantages for small values of the
throat radius.

\phantom{a}
\noindent
PAC numbers: 04.40.Nr, 04.20.Jb, 04.20.Dw
\end{abstract}

\section{Intoduction}
A new renormalizable gravity theory in four dimensions,
proposed by Ho\v{r}ava \cite{ HL}, may be regarded as a UV
complete candidate for general relativity.  At large
distances the theory reduces to Einstein gravity with a
non-vanishing cosmological constant in IR, but with improved
UV behavior.  As discussed in Ref. \cite{mP09}, from the
IR-modified Ho\v{r}ava action, which reduces to the
standard Einstein-Hilbert action in the IR limit, one
obtains the analogue of the standard spherically symmetric
Schwarzschild-(A)dS black-hole solution.

In this paper we employ such a class of black holes by starting
with two copies thereof and constructing a traversable thin-shell
wormhole by means of the cut-and-paste technique
\cite{Poisson1995}.  To compare and contrast the two
gravitational theories, we discuss various aspects of these
wormholes, such as the location of the event horizons, the
energy density of the thin shell, the violation of the
weak energy condition at the junction surface, the amount
of exotic matter required, and the equation of state.  Our
final topic is the question of stability to a linearized
spherically symmetric perturbation.

\section{Ho\v{r}ava-Lifshitz black holes}\label{S:HLBH}.

\noindent Consider the static and spherically symmetric line element
\begin{equation}\label{E:line1}
ds^{2}=-N(r)^2dt^{2}+\frac{dr^2}{f(r)}
    +r^{2}(d\theta^{2}+\text{sin}^{2}\theta\,d\phi^{2}).
\end{equation}
The modified Ho\v{r}ava action mentioned above is
\begin{multline}
  \mathcal{L}=\frac{\kappa^2\mu^2}{8(1-3\lambda)}
   \frac{N}{\sqrt{f}}\left[(2\lambda-1)\frac{(f-1)^2}{r^2}
    -2\lambda\frac{f-1}{r}f'+\frac{\lambda-1}{2}f'^2\right.\\
   \left.\phantom{\frac{N}{A^2}}-2(\omega-\Lambda_W)(1-f-rf')
    -3\Lambda^2_W r^2\right],
\end{multline}
where $\omega=8\mu^2(3\lambda-1)/\kappa^2$\,\,\cite{KS2009}.
Here $\lambda$ and $\kappa$ are dimensionless coupling
constants, while the constants $\mu$ and $\Lambda_W$ are
dimensionful.

The case $\lambda=1$, which reduces to the standard
Einstein-Hilbert action in the IR limit \cite{mP09}, yields

\begin{equation}\label{E:gtt}
N^2=f=  1+(w-\Lambda_W)r^2-
\sqrt{r[w(w-2\Lambda_W)r^3 + \gamma]}.
\end{equation}
Here $\gamma$ is an  integration constant.  By considering
$\gamma=-\alpha^2/\Lambda_W$ and $\omega=0$, this reduces
to the solution given by Lu, Mei, and Pope (LMP)
\cite{Lu2009}:
\begin{equation}
f=  1-\Lambda_Wr^2-
\frac{\alpha}{\sqrt{-\Lambda_W}}\sqrt{r}.
\end{equation}
If $\Lambda_W=0$ and $\gamma=4\omega M$ in Eq. (\ref{E:gtt}),
one obtains the Kehagias-Sfetsos (KS) black-hole solution
\cite{KS2009}:
\begin{equation}\label{E:KS}
f=  1+wr^2- wr^2\sqrt{1+ \frac{4M}{wr^3}}.
\end{equation}
Finally, since $\lambda=1$, we now have
\begin{equation}
  \omega=\frac{16\mu^2}{\kappa^2}.
\end{equation}

The Kehagias-Sfetsos solution is the only asymptotically flat
solution in the family of solutions (\ref{E:gtt}). We will
therefore use the Kehagias-Sfetsos solution for constructing
the thin-shell wormhole from  Ho\v{r}ava-Lifshitz black holes.
It is to be noted that there is an outer (event) horizon
$(r_+)$, and an inner (Cauchy) horizon $(r_-)$ of the
Kehagias-Sfetsos black-hole solution for $wM^2 > \frac{1}{2}$
at

\begin{equation}
r_\pm=  M\left[ 1\pm \sqrt{1-\frac{1}{2wM^2}}~\right],
\end{equation}
compared to $\sqrt{1-2M/r}$ for the Schwarzschild case, i.e.,
in Einstein gravity.

This spacetime comes back to the Schwarzschild black hole
for $ r \gg \left( \frac{M}{w}\right)^{1/3}$,
so that
\[
   f\approx 1-\frac{2M}{r}-\mathcal{O}(r^{-4}).
\]
Finally, observe that
\begin{multline}\label{E:eventhorizons}
 0 = r_-(Schwarzschild)<r_-(KS)=M\left[1-
\sqrt{1-\frac{1}{2wM^2}}~\right] \\ < r _+(KS)  =
M\left[ 1+\sqrt{1-\frac{1}{2wM^2}}~\right] < 2M=
r_+(Schwarzschild).
\end{multline}
The event horizons for various values of $\omega$ and their
relationship to the Schwarzschild case can be seen in Fig. 1.
\begin{figure}[tbp]
\begin{center}
\includegraphics[width=0.5\textwidth]{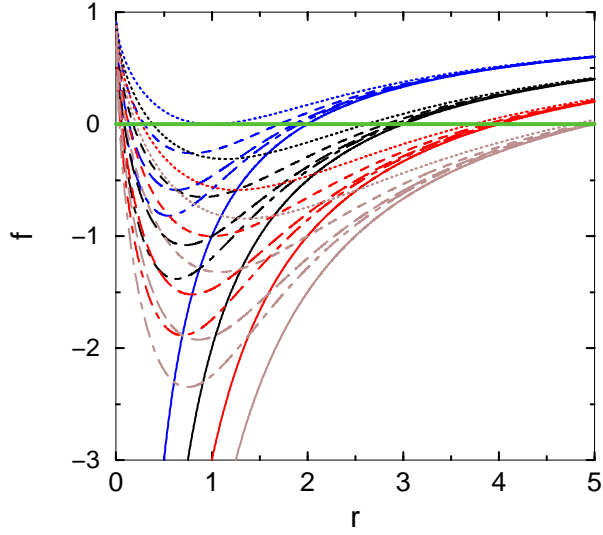}
\end{center}
\caption{The plot for $f$   using various parameters. The dotted,
dashed, long dashed, and chain curves correspond to $\omega=0.5,
1.0, 2.0, \text{and}\, 3.0$, respectively.  The thick solid lines
corresponds to the Schwarzschild case.  The blue, black, red and
brown colors represent  $M=1.0, 1.5, 2.0, \text{and}\, 2.5$,
respectively.}
\end{figure}

\section{Thin-shell wormholes from Ho\v{r}ava-Lifshitz black
holes }\label{S:thin-shell}

The mathematical construction of a thin-shell wormhole, when
first proposed by Visser \cite{Visser95}, used a flat Minkowski
space as a starting point.  The construction itself, however,
relies on a topological identification of two surfaces and is
not confined to a particular type of geometry.  So we can
safely start with two copies of a Kehagias-Sfetsos black hole
and remove from each the four-dimensional region
\[
  \Omega^\pm = \{r\leq a\,|\,a>r_h\}.
\]
We assume that $r_h=r_+$, the larger of the two radii.  The
construction proceeds by identifying (in the sense of topology)
the timelike hypersurfaces
\[
  \partial\Omega^\pm =\{r=a\,|\,a>r_h\},
\]
denoted by $\Sigma$.  Thanks to the asymptotic flatness referred
to above, the resulting manifold is geodesically complete and
consists of two asymptotically flat regions connected by a
throat, namely $r=a$.

Our main goal is to compare various characteristics of
Ho\v{r}ava-Lifshitz gravity and Einstein gravity, such as the
surface stress-energy tensor and the basic question of
stability.  A complete analysis would require, as always, a
detailed knowledge of the junction conditions, but the
traditional method, using the Lanczos equations, does not
automatically carry over to Kehagias-Sfetsos black holes
without an additional assumption.   So let us return to 
Eq. (\ref{E:eventhorizons}) and Fig. 1, neither of which 
depends on the junction conditions, to gain an overview: 
the effect of the convergence to the Schwarzschild case 
can be clearly seen for the event horizons.  But the 
convergence of the KS solution to the Schwarzschild 
solution is true in general and never depends on the 
junction conditions.  From this standpoint, all 
well-constructed plots can be expected to show certain 
trends that allow a comparison between the two 
gravitational theories.  Indeed, all the plots to 
follow show the gradual departure from the 
Schwarzschild limit as $a$ decreases.  The ability to 
make the relevant comparisons justifies the 
assumption that the thin-shell formalism is at least
\emph{qualitatively} acceptable, given that the use
of plots is essentially  qualitative in nature.
For example, Fig. 7 shows that the amount of exotic
matter required is \emph{less} for the KS case than
for the Schwarzschild case, which is indeed a
qualitative statement.  (This point will be
reiterated from time to time when making the
comparisons.)

A final observation in Fig. 1 is that the plot for 
$f$ starts taking on a very different form from that 
of the Schwarzschild case for small values of 
$r=a$, enough to cast some doubt on the validity.
This suspicion is readily confirmed by the
fact that in some cases the results are actually
unphysical for sufficiently small values of $a$; such
values may therefore be disregarded as meaningless.

With these caveats in mind, we will proceed with the
assumption that the induced metric on $\Sigma$ is
given by
\begin{equation}
               ds^2 =  - d\tau^2 + a(\tau)^2( d\theta^2 +
               \sin^2\theta \,d\phi^2),
\end{equation}
where $\tau$ is the proper time on the junction surface.  It
follows from the Lanczos equations
\cite{Poisson1995,Visser1989,Lobo2003,Lobo2004,Eiroa2004a,Eiroa2004b,Eiroa2005,
Thibeault2005,Lobo2005,Rahaman2006,
Eiroa2007,Rahaman2007a,Rahaman2007b,Rahaman2007c,Richarte2007,Lemos2008,Rahaman2008a,
Rahaman2008b,Eiroa2008a,Eiroa2008b, Rahaman2010a, Rahaman2010b},
\begin{equation*}\label{E:Lanczos}
  S^i_{\phantom{i}j}=-\frac{1}{8\pi}\left([K^i_{\phantom{i}j}]
   -\delta^i_{\phantom{i}j}[K]\right),
\end{equation*}
that the surface stress-energy tensor is
$S^i_{\phantom{i}j}=\text{diag}(-\sigma, p_{\theta},
 p_{\phi})$, where $\sigma$ is the surface energy density and
$p=p_{\theta}=p_{\phi}$ is the surface pressure. The Lanczos
equations then yield
\begin{equation*}\label{E:stress1}
  \sigma=-\frac{1}{4\pi}[K^\theta_{\phantom{\theta}\theta}]
\end{equation*}
and
\begin{equation*}\label{E:stress2}
  p=\frac{1}{8\pi}\left([K^\tau_{\phantom{\tau}\tau}]
    +[K^\theta_{\phantom{\theta}\theta}]\right).
\end{equation*}

Following Ref. \cite{Poisson1995}, a dynamic analysis can be
obtained by letting the radius $r=a$ be a function of time.
As a result,
\begin{equation}\label{E:s}
\sigma = - \frac{1}{2\pi a}\sqrt{f(a) + \dot{a}^2}
\end{equation}
and
\begin{equation}\label{E:p}
p_{\theta} = p_{\phi} = p =  -\frac{1}{2}\sigma + \frac{1}{8\pi
}\frac{2\ddot{a} + f^\prime(a) }{\sqrt{f(a) + \dot{a}^2}},
\end{equation}
where the overdot and prime denote, respectively, the
derivatives with respect to $\tau$ and $a$.  Here $ p$ and
$\sigma $ obey the conservation equation
\begin{equation}
               \frac {d}{d \tau} (\sigma a^2) + p \frac{d}{d \tau}(a^2)= 0
               \end{equation}
or
\begin{equation}\label{E:conservation}
               \dot{\sigma} + 2 \frac{\dot{a}}{a}( p + \sigma ) = 0.
               \end{equation}

For a static configuration of radius $a$, we need to assume
that $\dot{a} = 0 $ and $\ddot{a}= 0 $.  From  Eqs. (\ref{E:s})
and (\ref{E:p}) we have
\begin{equation}\label{E:sigmaKS}
\sigma_{KS} = - \frac{1}{2 \pi a }\sqrt{1+wa^2- wa^2\sqrt{1+
\frac{4M}{wa^3}}}
\end{equation}
in Ho\v{r}ava-Lifshitz gravity and
\begin{equation}\label{E:sigmaS}
\sigma_{Schwarzschild} = - \frac{1}{2 \pi a }\sqrt{1-
\frac{2M}{a}}
\end{equation}
in Einstein gravity.   Eq.(\ref{E:sigmaKS}) shows that
the energy density of the shell is negative in the KS
case, just as it is in the Schwarzschild case. We can
see from Fig. 2 that in the KS case the energy density
of the shell tends to be well below that of the
Schwarzschild case, as long as $a$ is not too small,
which is a considerable advantage from the standpoint
of wormhole design.  As $a$ gets large, however,
$\sigma_{KS}$ approaches
\begin{figure}[ptb]
\begin{center}
\vspace{0.5cm} \includegraphics[width=0.4\textwidth]{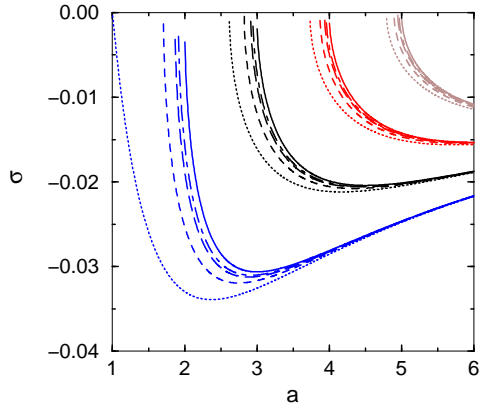}
\end{center}
\caption{The plot for $\sigma$ vs. $a$ for various parameters.  The
description of the curves is the same as in Fig. 1.
}%
\label{fig3}%
\end{figure}
$\sigma_{Schwarzschild}$, as expected from the discussion
in Sec. \ref{S:HLBH}.  Qualitatively, Fig. 3 shows a similar
advantagious behavior for the lateral pressure
\begin{equation}\label{E:pressureKS}
p_{KS}   = \frac{1}{8\pi} \left [
\frac{4aw-4aw\sqrt{1+ \frac{4M}{wa^3}}+\frac{6M}{a^2}
\frac{1}{\sqrt{1+4M/wa^3}}+
\frac{2}{a}}{\sqrt{1+wa^2-wa^2\sqrt{1+ \frac{4M}{wa^3}}}}\right]
\end{equation}
in Ho\v{r}ava-Lifshitz gravity and
\begin{equation}\label{E:pressureS}
p_{Schwarzschild} =  \frac{1}{4 \pi a }\left[
\frac{1-\frac{M}{a}}{ \sqrt{1- \frac{2M}{a}}}\right]
\end{equation}
in Einstein gravity.  Once again, the differences are greatest
for somewhat smaller values of $a$.  (Figs. 2 and 3 also show
the effect of having different event horizons.)
\begin{figure}[ptb]
\begin{center}
\vspace{0.5cm} \includegraphics[width=0.4\textwidth]{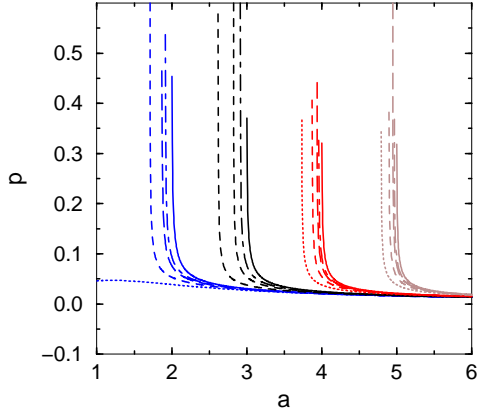}
\end{center}
\caption{The plot for the pressure $p$ vs. $a$ for various
parameters.  The description of the curves is the same as in Fig. 1.
}%
\label{fig3}%
\end{figure}

Since the shell is infinitely thin, the radial pressure
is zero.  So the matter on the shell violates both the weak
and null energy conditions.

\section{Equation of state}

Let us suppose that the EoS at the surface $\Sigma$ is $p=W
\sigma, W\equiv\text{constant}$. From Eqs. (\ref{E:sigmaKS})
-(\ref{E:pressureS}),
\begin{equation}
\frac{p_{KS}}{\sigma_{KS}}  = W_{KS} =  -\frac{a}{4} \left [
\frac{4aw-4aw\sqrt{1+ \frac{4M}{wa^3}}+\frac{6M}{a^2}
\frac{1}{\sqrt{1+4M/wa^3}}+ \frac{2}{a}}{
1+wa^2-wa^2\sqrt{1+ \frac{4M}{wa^3}}}\right]\end{equation}
in Ho\v{r}ava-Lifshitz gravity and
\begin{equation}
\frac{p_{Schwarzschild}}{\sigma_{Schwarzschild}}  =
W_{Schwarzschild} =  - \frac{1}{2 } \left [
\frac{1-\frac{M}{a}}{ 1- \frac{2M}{a}}\right] \end{equation}
in Einstein gravity.   Even though $W_{KS}>
W_{Schwarzschild}$,  $W_{KS}$  extends into the
phantom-energy regime (less than $-1$), but in both
cases, $W\rightarrow -1/2$ from below as $a\rightarrow \infty$,
as shown in Fig. 4.  The positive values are unphysical,
corresponding to locations well inside the event horizon.

\begin{figure}[ptb]
\begin{center}
\vspace{0.5cm} \includegraphics[width=0.4\textwidth]{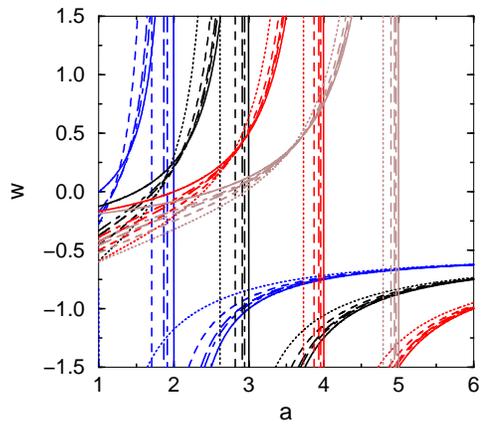}
\end{center}
\caption{The plot for $W$ vs. $a$ for various parameters.  The
description of the curves is the same as in Fig. 1.
}%
\label{fig3}%
\end{figure}

Another property worth checking is the traceless surface
stress-energy tensor $S^i_{\phantom{i}i}=0$, i.e.,
$-\sigma + 2p =0$.  The reason is that the Casimir effect with
a massless field is of the traceless type.  From this equation
we find that

\begin{multline}   C_{KS} \equiv  2 \left[ 1+wa^2- wa^2\sqrt{1+
\frac{4M}{wa^3}} ~\right]\\ + a  \left [ 4aw-4aw\sqrt{1+
\frac{4M}{wa^3}}+\frac{6M}{a^2} \frac{1}{\sqrt{1+
\frac{4M}{wa^3}}}+ \frac{2}{a}\right] = 0
\end{multline}
in Ho\v{r}ava-Lifshitz gravity and
\begin{equation} C_{Schwarzschild} \equiv  2 -\frac{3M}{a}  = 0 \end{equation}
in Einstein gravity.
\begin{figure}[ptb]
\begin{center}
\vspace{0.5cm} \includegraphics[width=0.4\textwidth]{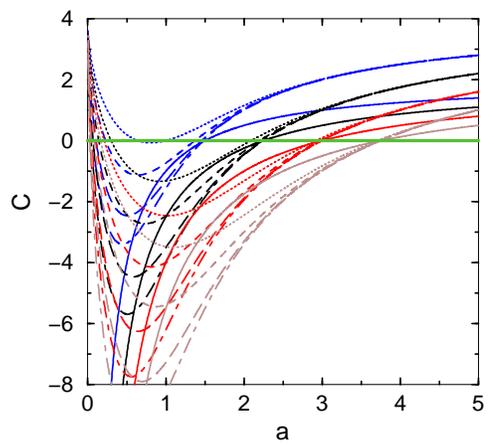}
\end{center}
\caption{The curves $C$ cut the $a$-axis at points less than the
horizon $(r_+)$.  The description of the curves is the same as in
Fig. 1.}%
\label{fig3}%
\end{figure}

We can see from Fig. 5 that the values of $a$ satisfying these
equations lie inside the event horizon, ensuring that this
situation cannot occur when dealing thin-shell wormholes.

\section{The gravitational field}
\noindent In this section we study the attractive or repulsive
nature of our wormhole.  To this end, we calculate the observer's
four-acceleration
\[ A^\mu = u^\mu_{; \nu} u^\nu, \]
where
\[u^{\nu} = \frac {d x^{\nu}}{d {\tau}} = \left( \frac{1}{\sqrt{f(r)}},
0,0,0\right).\] It follows from Eq. (\ref{E:gtt}) that the only nonzero
component is given by
\[ a^r = \Gamma^r_{tt} \left(\frac{dt}{d\tau}\right)^2
=   \alpha(r) ~, \] where
 \begin{equation}
 \alpha(r)_{KS} = wr -wr\sqrt{1+
\frac{4M}{wr^3}} +\frac{3M}{r^2} \frac{1}{\sqrt{1+
\frac{4M}{wr^3}}}
  \end{equation}
in Ho\v{r}ava-Lifshitz gravity and

\begin{equation} \alpha(r)_ {Schwarzschild}= \frac{M}{r^2} \end{equation}
in Einstein gravity.

A test particle   moving radially from rest obeys the geodesic
equation
\[ \frac{d^2r}{d\tau^2}= -\Gamma^r_{tt}\left(\frac{dt}{d\tau}\right)^2
   = -a^r.
\]
It is true in general that a wormhole is attractive whenever $
a^r>0 $ and repulsive whenever $a^r<0$.  According to Fig. 6,
the wormholes are attractive in both gravitational theories,
\begin{figure}[ptb]
\begin{center}
\vspace{0.5cm} \includegraphics[width=0.4\textwidth]{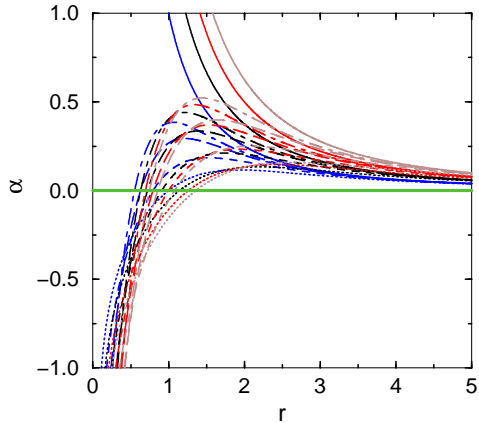}
\end{center}
\caption{The plot for $\alpha(r)$ vs. $a$ for various parameters.
The description of the curves is the same as in Fig. 1.
}%
\label{fig3}%
\end{figure}
since the negative values in the KS case are well within the
event horizon.  Continuing the qualitative theme for small values
of the throat radius, the acceleration is less for the KS case,
which is an advantage for a traveler through the wormhole.

\section{The total amount of exotic matter}
\noindent An important consideration in wormhole physics is
the total amount of exotic matter on the thin-shell.  This total
can be quantified by the integral
 \cite{Eiroa2005,Thibeault2005,Lobo2005,Rahaman2006,
Eiroa2007,Rahaman2007a,Rahaman2007b}
\begin{equation}
   \Omega_{\sigma}=\int [\rho+p]\sqrt{-g}d^3x.
\end{equation}
By introducing the radial coordinate $R=r-a$, we get
\[
 \Omega_{\sigma}=\int^{2\pi}_0\int^{\pi}_0\int^{\infty}_{-\infty}
     [\rho+p]\sqrt{-g}\,dR\,d\theta\,d\phi.
\]
Since the shell is infinitely thin, it does not exert any radial
pressure, while $\rho=\delta(R)\sigma(a)$.  So
\begin{multline}
 \Omega_{\sigma~ KS}=\int^{2\pi}_0\int^{\pi}_0\left.[\rho\sqrt{-g}]
   \right|_{r=a}d\theta\,d\phi=4\pi a^2\sigma(a)\\
  =-2a \sqrt{1+wa^2- wa^2\sqrt{1+
\frac{4M}{wa^3}}}
\end{multline}
in Ho\v{r}ava-Lifshitz gravity and
\begin{equation}  \Omega_{\sigma ~Schwarzschild }= -2 a   \sqrt{1-
\frac{2M}{a}} \end{equation}
in Einstein gravity.  According to Fig. 7, for small $a$ the amount
of exotic matter required is considerably less and may even be 
very much less for the KS than for the Schwarzschild case, which 
may be viewed as the most
\begin{figure}[ptb]
\begin{center}
\vspace{0.5cm}
\includegraphics[width=0.4\textwidth]{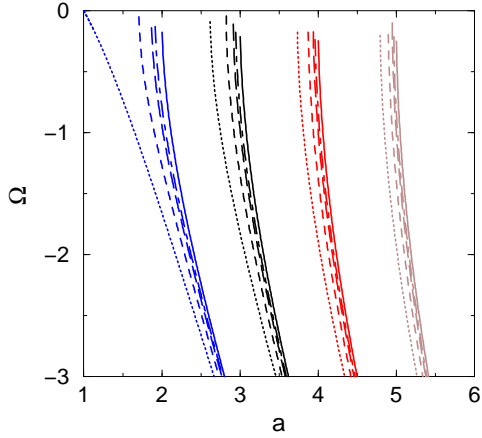}
\end{center}
\caption{The plot for $\Omega_{\sigma ~KS}$ and
$\Omega_{\sigma ~Schwarzschild}$ for various parameters.  The
description of the curves is the same as in Fig. 1.
}%
\label{fig3}%
\end{figure}
important difference between the two gravitational theories
from the standpoint of wormhole design.  (Once again, both
Figs. 6 and 7 suggest that $a$ must not be too small.)

\section{Linearized stability}\noindent
In this section we turn to the question of stability of the
wormhole under small perturbations around a static solution
$a=a_0$.  It must be emphasized, however, that the analysis
is subject to the same limitations already discussed in
Sec. \ref{S:thin-shell} and, as will be seen, clearly
parallels our earlier findings.

Rearranging Eq. (\ref{E:s}), we obtain the equation of motion
of the thin shell:
\begin{equation}\label{E:motion}
\dot{a}^2 + V(a)= 0,
\end{equation}
where the potential $V(a)$ is defined  as
\begin{equation}\label{E:potential}
V(a) =  f(a) - \left[2\pi a \sigma(a)\right]^2.
\end{equation}
Following Ref. \cite{Poisson1995}, if we expand $V(a)$ around
$a_0$, we obtain
\begin{eqnarray}
V(a) &=&  V(a_0) + V^\prime(a_0) ( a - a_0) +
\frac{1}{2} V^{\prime\prime}(a_0) ( a - a_0)^2  \nonumber \\
&\;& + O\left[( a - a_0)^3\right],
\end{eqnarray}
where the prime denotes the derivative with respect to $a$.
Linearizing around $ a = a_0 $ requires that $ V(a_0) =0 $
and $ V^\prime(a_0)= 0 $.  The configuration will be in stable
equilibrium if $ V^{\prime\prime}(a_0)> 0 $.  As suggested in
Ref. \cite{Poisson1995}, the subsequent analysis will make use
of a parameter $\beta$, which is usually interpreted as the
subluminal speed of sound and is given by the relation
\[
 \beta^2(\sigma) =\left. \frac{ \partial
p}{\partial \sigma}\right\vert_\sigma.
\]
To determine the conditions that yield $V''(a_0)>0$, it is
convenient to start with Eq. (\ref{E:conservation}) and first
deduce that $(a\sigma)'=-(\sigma+2p)$.  Also,
\begin{equation}
  (a\sigma)''=-(\sigma'+2p')=-\sigma'\left(1+2
     \frac{\partial p}{\partial\sigma}\right)\\
   =2\left(1+2\frac{\partial p}{\partial\sigma}\right)
    \frac{\sigma+p}{a}=2(1+2\beta^2)\frac{\sigma+p}{a}.
\end{equation}
Returning to Eq. (\ref{E:potential}), we now readily obtain
\[
  V'(a)=f'(a)+8\pi^2a\sigma(\sigma+2p)
\]
and
\begin{equation*}
   V''(a)=f''(a)-8\pi^2(\sigma+2p)^2
   -8\pi^2[2\sigma(1+\beta^2)(\sigma+p)].
\end{equation*}
By direct computation we can now check the required
conditions $V(a_0)=0$ and $V'(a_0)=0$.  The stability
condition $V''(a_0)>0$ then yields the intermediate result
\begin{equation}\label{E:intermediate}
  2\sigma(\sigma+p)(1+2\beta^2)<\frac{f''(a_0)}{8\pi^2}
    -(\sigma+2p)^2.
\end{equation}
Recall that $\sigma$ is negative.  If $\sigma+p$ is also
negative, then the sense of the inequality is retained
in the next step.  If $\sigma+p$ is positive, then the
sense of the inequality is reversed.  (See Fig. 8.)  In
the former case we get
\begin{figure}[tbp]
\begin{center}
\includegraphics[width=0.5\textwidth]{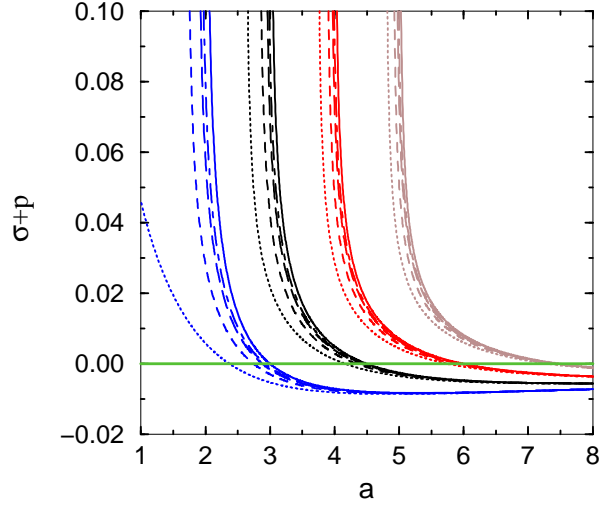}
\end{center}
\caption{The plot for $\sigma+p$ vs. $a$ for various parameters.
The description of the curves is the same as in Fig. 1.
}
\end{figure}

\begin{equation}\label{E:less}
  \beta^2<\frac{\frac{f''(a_0)}{8\pi^2}-(\sigma+2p)^2
      -2\sigma(\sigma+p)}
   {2[2\sigma(\sigma+p)]}
\end{equation}
and in the latter case
\begin{equation}\label{E:more}
  \beta^2>\frac{\frac{f''(a_0)}{8\pi^2}-(\sigma+2p)^2
      -2\sigma(\sigma+p)}
   {2[2\sigma(\sigma+p)]}.
\end{equation}
For these inequalities,
\[ f(a_0) = 1+wa_0^2- wa_0^2\sqrt{1+ \frac{4M}{wa_0^3}},\]
\[ f^\prime(a_0) = 2wa_0- 2wa_0\sqrt{1+ \frac{4M}{wa_0^3}}
   +\frac{6M}{a_0^2} \frac{1}{\sqrt{1+ \frac{4M}{wa_0^3}}},\]
and
\[ f^{\prime\prime}(a_0) = 2w- 2w\sqrt{1+ \frac{4M}{wa_0^3}}
    +\frac{36M^2}{wa_0^6} \frac{1}
{\left(1+ \frac{4M}{wa_0^3}\right)^{3/2}}\]
in the KS case and
\[ f(a_0) = 1-\frac{2M}{a_0},\]
\[ f^\prime(a_0) = \frac{2M}{a_0^2},\]
and
\[ f^{\prime\prime}(a_0) =-\frac{4M}{a_0^3}\]
 in the Schwarzschild case.
\begin{figure}[ptb]
\begin{center}
\vspace{0.5cm} \includegraphics[width=0.4\textwidth]{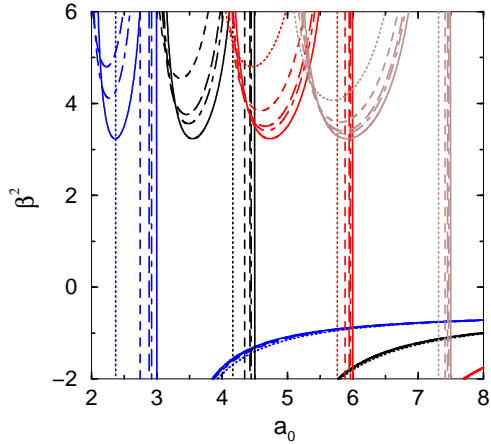}
\end{center}
\caption{The plot for $\beta^2$ vs. $a_0$ for
various parameters.  The description of the curves is the same as
in Fig. 1.
}%
\label{fig3}%
\end{figure}

Fig. 9 shows the region of stability for both the KS and
Schwarzschild cases.  According to Eqs. (\ref{E:less}) and
(\ref{E:more}) and Fig. 8, the respective regions of stability
are below the curves on the right and above the curves on the left.

Under ordinary circumstances, $\beta$ represents the velocity
of sound, since $\beta^2=\partial p/\partial\sigma$ is the square of
the rate of change of distance with respect to time.  So, ordinarily,
$0<\beta^2\le 1$.  For the Schwarzschild case \cite{Poisson1995},
the minimum value is $3/2+\sqrt{3}$ .  According to Fig. 9, for
the KS case, the region of stability involves values of $\beta^2$
that are even larger.  (Of course, for the region on the right,
$\beta^2$ is negative.)  So by this criterion, there are no
stable solutions for either type of wormhole.

All these interpretations are valid as long as $\beta$ is indeed
the speed of sound.  As discussed extensively in Ref.
\cite{Poisson1995}, this assumption should be treated with caution
when dealing with exotic matter.  In this paper we have the
additional complication that the KS case is based on a
nonrelavistic gravitational theory.  So we cannot exclude the
possibility that $\beta^2$ is just a convenient parameter.

\section{Conclusion}
This paper discusses a new class of thin-shell wormholes from
black holes in Ho\v{r}ava-Lifshitz gravity by employing the
asymptotically flat Kehagias-Sfetsos (KS) solution with various
values of the coupling constant $\omega$ and the mass $M$.  In
all cases the radius of the outer event horizon in the KS case
 turns out to  be less than that in the Schwarzschild case.
A wormhole designer from an advanced civilization would find
several advantages in the KS case over the Schwarzschild case:
for small values of the throat radius $r=a$, the negative
energy density of the thin shell is well below that in the
Schwarzschild case.  The lateral pressure is also less.
A particularly interesting finding is that for small
values of $a$, the amount of exotic matter required can be
much less for a KS than for a Schwarzschild wormhole, a
considerable advantage given the problematical nature of
exotic matter.  As $a$ gets large, however, the properties
of the KS wormhole approach those of the Schwarzschild
wormhole.  A similar conclusion holds for the equation of
state, assumed to be of the form $p=W\sigma$:
while $W_{KS}>W_{Schwarzschild}$ for small $a$, in
both gravitational theories $W\rightarrow -1/2$ as
$a\rightarrow\infty$.  While both types of wormholes are
attractive, the acceleration toward the center is less
for the KS case than for the Schwarzschild case, a
particular advantage for a traveler.  The final topic
is a discussion of the stability to linearized radial
perturbations in terms of a parameter $\beta$, normally
interpreted as the speed of sound, but which may be just
a convenient parameter.  It was found that stable
solutions exist in the KS case for values of $\beta$
that are similar to the values in the Schwarzschild
case.
\begin{center}
  \textbf{Acknowledgment}

\phantom{a}
FR is grateful for the financial support from UGC, Govt. of India.
\end{center}


\begin{thebibliography}{99}


\bibitem{HL}  P. Ho\v{r}ava, arXiv:0811.2217 [hep-th];
P. Ho\v{r}ava, JHEP {\bf903}, 20 (2009) [arXiv:0812.4287
[hep-th]]; P. Ho\v{r}ava, Phys. Rev. D {\bf79}, 084008 (2009)
[arXiv:0901.3775 [hep-th]]; P. Ho\v{r}ava, arXiv:0902.3657
[hep-th].
\bibitem{mP09}Mu-in Park, JHEP \textbf{09}, 123 (2009).
\bibitem{Poisson1995} E. Poisson and M. Visser, Phys. Rev. D {\bf 52}, 7318 (1995).
\bibitem{KS2009} A. Kehagias and K. Sfetsos, Phys. Lett. B {\bf678}, 123 (2009).
\bibitem{Lu2009}H. Lu, J. Mei and C. N. Pope, arXiv:0904.1595 [hep-th].
\bibitem{Visser1989} M. Visser, Nucl. Phys. B 328, 203 (1989).
\bibitem{Lobo2003} F.S.N. Lobo and P. Crawford, Class. Quant. Grav. {\bf 21}, 391 (2004).
\bibitem{Lobo2004} F.S.N. Lobo, Class. Quant. Grav. {\bf 21}, 4811 (2004).
\bibitem{Eiroa2004a} E.F. Eiroa and G. Romero, Gen. Rel. Grav. {\bf 36}, 651 (2004).
\bibitem{Eiroa2004b} E.F. Eiroa and C. Simeone, Phys. Rev. D {\bf 70}, 044008 (2004).
\bibitem{Eiroa2005} E.F. Eiroa and C. Simeone, Phys. Rev. D {\bf 71}, 127501 (2005).
\bibitem{Thibeault2005} M. Thibeault, C. Simeone, and E.F. Eiroa, Gen. Rel. Grav.
    {\bf 38}, 1593 (2006).
\bibitem{Lobo2005} F.S.N. Lobo, Phys. Rev. D {\bf 71}, 124022 (2005).
\bibitem{Rahaman2006} F. Rahaman et al., Gen. Rel. Grav. {\bf 38}, 1687 (2006).
\bibitem{Eiroa2007}  E.F. Eiroa and C. Simeone, Phys. Rev. D {\bf 76}, 024021 (2007).
\bibitem{Rahaman2007a} F. Rahaman et al., Int. J. Mod. Phys. D {\bf 16}, 1669 (2007).
\bibitem{Rahaman2007b} F. Rahaman et al., Gen. Rel. Grav. {\bf 39}, 945 (2007).
\bibitem{Rahaman2007c} F. Rahaman et al., Chin. J. Phys.
 {\bf45}, 518 (2007) arXiv:0705.0740 [gr-qc]
\bibitem{Richarte2007} M. G. Richarte and C. Simeone, Phys. Rev. D {\bf76},
    087502 (2007).
\bibitem{Lemos2008} J. P. S. Lemos and F.S.N. Lobo, Phys. Rev D {\bf78},
    044030 (2008).
\bibitem{Rahaman2008a} F. Rahaman et al., Acta Phys. Polon. B {\bf40}, 1575 ( 2009 )
    arXiv: gr-qc/0804.3852.
\bibitem{Rahaman2008b} F. Rahaman et al., Mod. Phys. Lett. A {\bf24}, 53 (2009) arXiv:
gr-qc/0806.1391.
\bibitem{Eiroa2008a} E.F. Eiroa, Phys. Rev. D {\bf 78}, 024018 (2008).
\bibitem{Eiroa2008b} E.F. Eiroa, M.G. Richarte, and C. Simeone, Phys. Lett. A {\bf 373}
    1 (2008).
  \bibitem{Rahaman2010a} F. Rahaman, K A Rahman, Sk.A Rakib,   Peter K.F.
  Kuhfittig, Int. J. Theor. Phys. {\bf 49}, 2364 (2010).
e-Print: arXiv:0909.1071 [gr-qc]
 \bibitem{Rahaman2010b} A.A. Usmani, Z. Hasan , F. Rahaman, Sk.A. Rakib,  , Saibal Ray , Peter K.F. Kuhfittig,
Gen. Relativ. Gravit.  {\bf 42},  2901 (2010) e-Print:
arXiv:1001.1415 [gr-qc]
\bibitem{Visser95}M. Visser, \emph{Lorentzian wormholes$-$from Einstein to Hawking} (American
   Institute of Physics, New York, 1995).

\end{thebibliography}
\end{document}